\documentclass[aps,amsfonts,preprint]{revtex4}
\usepackage{epsfig,amsmath,amssymb,bm,epsf,graphics,psfrag,color,ulem}

\def\changed{}

\def\Longarrow{\protect\@lra}
\def\@lra{\relbar\joinrel\relbar\joinrel\relbar\joinrel%
          \relbar\joinrel\rightarrow}

\def\veps{\varepsilon}

\begin{document}
\title{Revisiting the critical velocity of a clean one-dimensional superconductor }

\author{Tzu-Chieh Wei}
\affiliation{Institute for Quantum Computing and Department of Physics and
Astronomy, University of Waterloo, Waterloo, ON N2L 3G1, Canada}
\altaffiliation[Present address:\,]{Department of Physics and Astronomy,
University of British Columbia, Vancouver, BC V6T 1Z1, Canada}
\author{Paul M. Goldbart}
\affiliation{Department of Physics, Institute for Condensed Matter Theory,
 and Federick Seitz Materials Research Laboratory, University of Illinois at Urbana-Champaign, Urbana,
Illinois 61801, U.S.A.}

\date{April 15, 2009}

\begin{abstract}
We revisit the problem of the critical velocity of a clean one-dimensional
superconductor. {\changed At the level of mean-field theory}, we find that the
zero-temperature value of the critical velocity---the uniform velocity of the
superfluid condensate at which the superconducting state becomes unstable---is
a factor of $\sqrt{2}$ smaller than the Landau critical velocity. This is in
contrast to a prior finding, which held that the critical velocity is equal to
the Landau critical velocity. The smaller value of the critical velocity,
which our analysis yields, is the result of a pre-emptive
Clogston-Chandrasekhar--like discontinuous phase transition, and is an analog
of the threshold value of the uniform exchange-field of a superconductor,
previously investigated by Sarma and by Maki and Tsuneto. We also consider the
impact of nonzero temperature, study critical currents, and examine
metastability and its limits in the temperature versus flow-velocity phase
diagram. In addition, we comment on the effects of electron scattering by
impurities.
\end{abstract}
\pacs{74.25.Sv, 74.62.-c, 74.25.Dw} \maketitle
\section{Introduction}\label{sec:Intro}
The Landau criterion~\cite{Landau} concerns the threshold velocity $v_L$ of an
 obstacle moving through stationary superfluids at zero temperature, beyond
which excitations are created and superfluidity is lost. By Galilean
invariance, the criterion also applies to the threshold velocity $v_L$ of a
uniformly flowing superfluid; see, e.g., Ref.~\cite{Varoquaux06}. The value of
the threshold velocity $v_L$ is determined by the following formula
\begin{equation}
v_L=\min_{p}\left(\frac{E_p}{p}\right),
\end{equation}
where $E_p$ is the excitation spectrum. For superconductors the Landau
criterion gives $v_L=\Delta_0/p_F$, where $2\Delta_0$ is the pairing gap at
zero temperature and zero flow, and $p_F=\hbar k_F$ is the Fermi momentum of
the entities that are paired (with $k_F$ being the associated angular
wavenumber). However, it was found by Rogers~\cite{Rogers,Bardeen,Zagoskin}
that in three spatial dimensions the critical velocity $v_c$---the uniform
velocity of the superfluid condensate at which the superconducting state
becomes unstable---of a clean superconductor exceeds $v_L$. Furthermore, as
discussed, e.g., in Ref~\cite{Zagoskin}, for superflow velocities $v$ in the
range $v_L\le v\le v_c$, gapless excitations occur in clean superconductors.
Moreover, the ratio $v_c^{\rm (3D)} /v_L$ has been found to be $e/2$
($\approx\, 1.359$). In contrast, it was found~\cite{Zagoskin} that  in two
dimensions $v_c^{\rm (2D)}/v_L=1$. As for the case of one dimension, Bagwell
reported~\cite{Bagwell} that $v_c^{\rm (1D)}/v_L=1$, as found in two
dimensions. In contrast to the case of three dimensions, in neither one nor
two dimensions is gapless superconductivity predicted to occur in the presence
of flow.

  In this Paper, we re-analyze the critical velocity of a clean superconductor
in one spatial dimension {\changed via a mean-field treatment}, and obtain the
dependence of this velocity on the temperature $T$, along with the associated
temperature-velocity phase diagram. In particular, we find that at zero
temperature the critical velocity is smaller by a factor of $\sqrt{2}$ than
the Landau critical velocity $v_L$. Even though for $v=v_L/\sqrt{2}$ a gap in
the quasiparticle excitation spectrum remains, the superconducting state first
becomes unstable there. This is in contrast to a previous
report~\cite{Bagwell}, which held that the critical velocity is $v_L$, i.e.,
the velocity at which the gap in the quasiparticle spectrum closes and
excitations proliferate so as to destroy superconductivity. The smaller value
of the critical velocity, which we obtain here, is the result of a pre-emptive
Clogston-Chandrasekhar--like discontinuous phase
transition~\cite{ClogstonChandrasekhar}, and is analogous to the threshold
uniform exchange-field in a superconductor, previously investigated by
Sarma~\cite{Sarma} and by Maki and Tsuneto~\cite{MakiTsuneto}.
 At low temperatures (i.e., for $T < T^*\approx 0.556\, T_c^0$, where $T_c^0$ is the critical
temperature in the absence of flow), the transition from the superconducting
to the normal state, occurring
 due to the presence of flow, remains
discontinuous. In contrast, for $T\ge T^*$ the transition is continuous, i.e.,
it is associated with a continuously vanishing order parameter, e.g., as $T$
approaches the flow-velocity--dependent transition temperature $T_c(v)$ from
below.

From the experimental standpoint, it is usually more convenient to control the
condensate {\it current\/} rather than its {\it velocity}. (An exception is
provided by a closed loop threaded by magnetic flux.)  The more widely
appropriate physical observable is thus the critical { current\/}, i.e., the
maximum equilibrium current that a superconductor can sustain, and this is
what experiments on superconductors frequently measure. It should, however, be
remarked that recent experiments on trapped condensates of atomic gases make
it possible to probe the critical velocity directly~\cite{KetterleGroup}.

Langer and Fisher~\cite{LangerFisher67} introduced a fresh perspective on the
issue of the critical velocity of a sueprfluid. They explained that---as a
matter of principle---superflow is inherently unstable, even at velocities
below those that allow quasiparticle excitations to proliferate. This
instability results from the possible occurrence of intrinsic, topologically
allowed, collective excitation events, in which, e.g., a vortex ring nucleates
locally, traverses an Arrhenius energy barrier, and grows without bound, thus
eradicating a quantum of flow velocity.
 In the case of narrow
channels, such as the one-dimensional superconductors discussed here, the
fluctuations take the form of phase slip events, which can be either
thermal~\cite{LA} or quantum~\cite{QPS}. But, before one takes into account
the effects of fluctuations, it is important to understand and clarify the
behavior at the mean-field level, which is mainly what we consider in the
present Paper. In other words, the work reported here assumes that the rate of
such toplogical fluctuations is negligibly small, so that our
critical-velocity results, established on the basis of thermodynamic stability
and the possibility of quasiparticle production, retain a use.
{\changed In
the present Paper we focus on issues of phases of thermodynamic equilibrium,
competitions between them, and metastability; we do not attempt to address
issues of kinetics, such as the rates at which phase transitions proceed and
metastable states evolve into stable ones.}

The remainder of the Paper is organized as follows. In Sec.~\ref{sec:Hami} we
diagonalize the Hamiltonian of a superconductor in the presence of  flow, and
in Sec.~\ref{sec:FreeEn}
 we calculate the free energy of this system. In
Sec.~\ref{sec:OrderPara} we use the self-consistency of the order parameter to
calculate  the dependence of the order parameter on the flow velocity at
various temperatures. In Sec.~\ref{sec:TransitionT} we calculate the
dependence of the transition temperature on the flow velocity, assuming  the
transition to be continuous, i.e., in the limit of linear instability. Double
solutions for the transition temperature turn out to exist for larger values
of the flow velocity. We resolve this issue in Sec.~\ref{sec:Phase} by
identifying the globally stable solution, i.e., the one that corresponds to
the lower free energy. {\changed In Sec.~\ref{sec:Super} we address the
existence of metastable solutions, and discuss the extent to which these
provide a superflow-based analog of the
 phenomena of supercooling and superheating.} We also obtain the analog of the
superheating limit, together with the equilibrium phase boundary and the
supercooling limit. Having ascertained the true, equilibrium,
velocity-dependent order parameter, we calculate in Sec.~\ref{sec:current}
 the dependence of the supercurrent on the flow velocity, and hence determine the critical current and
 superfluid density. In Sec.~\ref{sec:Disorder} we briefly discuss the effect
of disorder on the nature of the transition. In Sec.~\ref{sec:higherD} we
contrast the critical-velocity results for one dimension with those for two
and three dimensions. We conclude in Sec.~\ref{sec:Conclude}.
\section{Hamiltonian}\label{sec:Hami}
We shall use a microscopic approach to discuss an effectively one-dimensional
 superconducting system, in which there is spatially uniform flow at velocity $v\equiv
 q/m$.
 Here, $m$ is the
electron mass and $q$ is the corresponding momentum associated with the flow
(or equivalently, the angular wavenumber, as we shall set $\hbar=1$). (What we
mean by ``effectively one-dimensional'' is that the transverse dimensions are
much smaller than the zero-temperature coherence length $\xi_0$.) We begin by
writing down a Hamiltonian that is equivalent to Eq.~(III.4) of
Rogers~\cite{Rogers} and Eq.~(1) of Nozi\`eres and
Schmitt-Rink~\cite{NozieresSchmitt-Rink}:
\begin{equation}
H=\sum_{k,\sigma}\left(\frac{(k+q)^2}{2m}-\mu\right) c_{k+q,\sigma}^\dagger\,
c_{k+q,\sigma} +\sum_{k_1,k_2}   V_{k_1,k_2}\, c_{k_1+q,\uparrow}^\dagger\,
c_{-k_1+q,\downarrow}^\dagger\, c_{-k_2+q,\downarrow}\, c_{k_2+q,\uparrow}\,,
\end{equation}
where $c_{k\sigma}$ and $c^\dagger_{k\sigma}$ are respectively the
annihilation and creation operators of electrons of spin-projection
$\sigma=\uparrow\,{\rm or}\,\downarrow$ at (one-dimensional) momentum $k$,
$V_{kk'}$ is the BCS pairing potential~\cite{BCS}, and $\mu$ is the chemical
potential. {\changed Following Rogers~\cite{Rogers}, we assume that $V_{kk'}$
depends on the difference between $k$ and $k'$, and hence is independent of
$q$. To justify our ignoring any $q$ dependence, we note that the critical
momentum $q$ is no bigger than roughly $m v_L$, and this is roughly
$(a/\xi_0)k_F$, i.e., several orders of magnitude smaller than $k_F$, where
$a$ is the atomic spacing and $\xi_0$ is the zero-temperature superconducting
coherence length.}
 This form of Hamiltonian anticipates that any Cooper pairs are formed via the pairing
 of electrons in states
$(k+q,\uparrow)$ and  $(-k+q,\downarrow)$, and that the resulting pairs have
center-of-mass momentum $2q$.

Next, we make the mean-field approximation and adopt the BCS form for
$V_{kk'}$~\cite{BCS}, thus arriving at the Hamiltonian
\begin{eqnarray}
H&=&\frac{|\Delta|^2}{g}+\sum_{k} \left(\varepsilon_{q}(k)\,
c_{k+q,\uparrow}^\dagger\,
c_{k+q,\uparrow}+\varepsilon_{q}(-k)\,c_{-k+q,\downarrow}^\dagger\,
c_{-k+q,\downarrow} \right.\nonumber\\
&&\quad\quad\left. - \big(\Delta^*\,  c_{-k+q,\downarrow} \, c_{k+q,\uparrow}+
\Delta\, c_{k+q,\uparrow}^\dagger\, c_{-k+q,\downarrow}^\dagger\big)\right),
\end{eqnarray}
in which $g=|V_{kk'}|$ is the magnitude of  $V_{kk'}$ in the momentum range
for which the pairing potential is nonzero, $\veps_{q}(\pm k)\equiv (q\pm
k)^2/2m-\mu$, and $\Delta=-\sum_{k'} V_{k,k'}\langle
c_{-k'+q,\downarrow}\,c_{k'+q,\uparrow}\rangle$ is the self-consistency
condition on the order parameter (and similarly for $\Delta^*$). This
mean-field Hamiltonian can be solved via the Bogoliubov-Valatin transformation
  \begin{equation}
  \label{eqn:BogoVale}
  c_{k+q,\uparrow}=u_k \gamma_{1;k} +v_k \gamma_{2;k}^\dagger
  \quad {\rm and}\quad c_{-k+q,\downarrow}=u_k \gamma_{2;k} - v_k \gamma_{1;k}^\dagger\,,
  \end{equation}
  with
  \begin{equation}
  u_k^2=\frac{1}{2}\left(1+\frac{\overline{\veps}_k}{E_k}\right)
  \quad {\rm and}\quad
   v_k^2=\frac{1}{2}\left(1-\frac{\overline{\veps}_k}{E_k}\right),
  \end{equation}
  where, for convenience, we have made the definitions
   $\overline{\veps}_k\equiv \big(\veps_q(k)+\veps_q(-k)\big)/2=k^2/2m-\big(\mu-q^2/2m\big)$
   and $E_k\equiv\sqrt{\overline{\veps}_k^2+|\Delta|^2}$. We choose $\Delta$ to be real and non-negative,
  and thus may drop the absolute value on $\Delta$ in the definition of $E_k$.
With this procedure, the Hamiltonian  becomes diagonal:
  \begin{equation}
  H=\frac{\Delta^2}{g}+\sum_k\left( \big(E_k+kv\big)\gamma_{1;k}^\dagger\gamma_{1;k}+
  \big(E_k-kv\big)\gamma_{2;k}^\dagger\gamma_{2;k}\right)+\sum_k(\overline{\veps}_k-E_k).
  \end{equation}
  This expression also holds for higher dimensions, provided we
  replace $k$ by $\vec{k}$, $v$ by $\vec{v}$, and $kv$ by $\vec{k}\cdot\vec{v}$ .
\section{Free energy}\label{sec:FreeEn}
As the Hamiltonian has been diagonalized, we can readily evaluate the
Helmholtz free energy $F$ of the system:
\begin{equation}\label{eqn:FreeEn0}
F\equiv\langle H\rangle -T S=\langle H\rangle+T\sum_k \Big[f_{1;k}\ln
f_{1;k}+(1-f_{1;k})\ln (1-f_{1;k})+ f_{2;k}\ln f_{2;k}+(1-f_{2;k})\ln
(1-f_{2;k}) \Big],
\end{equation}
where $f_1$ and $f_2$ are Fermi distribution functions, defined via
\begin{equation}
\label{eqn:f12}
 f_{1;k}\equiv \langle
\gamma^\dagger_{1;k}\gamma_{1;k}\rangle=\frac{1}{e^{\beta(E_k+kv)}+1}  \quad
{\rm and}\quad f_{2;k}\equiv \langle
\gamma^\dagger_{2;k}\gamma_{2;k}\rangle=\frac{1}{e^{\beta(E_k-kv)}+1},
\end{equation}
$\langle\cdots\rangle$ indicates an average, weighted by the equilibrium
density matrix, and $\beta\equiv 1/T$ is the inverse temperature (with
Boltzmann's constant set to unity). We therefore arrive at the result:
\begin{equation}
\label{eqn:FreeEn} F=\frac{\Delta^2}{g}+\sum_k \overline{\veps}_k-T\sum_k
\ln\big(2\cosh\beta E_k+2\cosh\beta{kv}\big).
\end{equation}
{\changed One may equally well calculate the partition function $Z$, so as to
obtain the free energy via $F=-k_B T\ln (Z)$}. The order parameter is to be
determined self-consistently, via
\begin{equation}
\label{eqn:self_con1} \Delta=g\sum_k\langle
c_{-k+q,\downarrow}\,c_{k+q,\uparrow}\rangle=g\sum_k\frac{\Delta}{2E_k}(1-f_{1;k}-f_{2;k}).
\end{equation}
We note that because the momentum sum runs over both positive and negative
values one, can safely replace $f_{1;k}$ by $f_{2;k}$ (or vice versa) in this
equation, and thus arrives at the result:
\begin{equation}
\label{eqn:self_con2} \Delta=g\sum_k\frac{\Delta}{2E_k}(1-2f_{2;k}),
\end{equation}
which is identical to Eq.~(21) of Ref.~\cite{Bagwell}. This equation can also
be derived by  demanding stationarity of the free energy, i.e., $\delta
F/\delta \Delta=0$. The above results, Eqs.~(\ref{eqn:FreeEn0})
to~(\ref{eqn:self_con2}) hold for two- and three-dimensional systems as well,
provided we
  replace $k$ by $\vec{k}$, $v$ by $\vec{v}$, and $kv$ by $\vec{k}\cdot\vec{v}$, but in the following we shall
  focus mainly on the case of one dimension.
\section{Order parameter}\label{sec:OrderPara}
From Eq.~(\ref{eqn:self_con1}), or equivalently Eq.~(\ref{eqn:self_con2}), we
can determine the values of the order parameter at any temperature and
superflow velocity. Of course, there is always the trivial solution
$\Delta=0$, which corresponds to the normal state. However, in this section
our focus is on nontrivial solutions.
\begin{figure}
\vspace{0.5cm} \centerline{ \rotatebox{0}{
        \epsfxsize=7.0cm
        \epsfbox{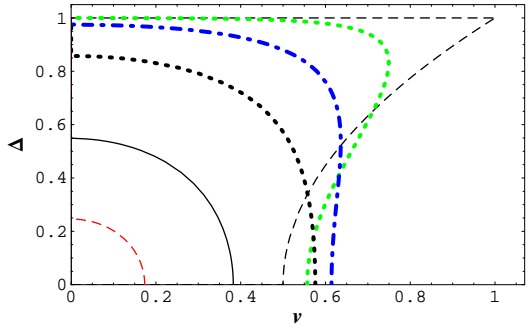}
} }
 \caption{Self-consistent solutions for the order parameter $\Delta$  (normalized to the zero-temperature, zero-flow value $\Delta_0$) vs.~superfluid velocity $v$ (in units of $v_L$) at various
 temperatures $T=(0, 0.223, 0.445, 0.668, 0.890, 0.980)\, T_c^0$ (from top-right to bottom-left). Note the multivaluedness of $\Delta$, which
 occurs for the lowest three values of the temperatures. {\changed It turns out that the
 lower branches of the order parameter  have the maximum in the free energy compared to the
 upper branches and the trivial $\Delta=0$ solution.}} \label{fig:1}
\end{figure}
\subsection{Zero temperature}
Let us first examine the limit of zero temperature. In this limit, the
self-consistency condition~(\ref{eqn:self_con2}) becomes
\begin{equation}
\Delta(T,v)\big|_{T=0}=
g\sum_{k>0}\frac{\Delta}{E_k}\Big(1-\Theta(kv-E_k)\Big),
\end{equation}
where $\Theta(x)$ is the Heaviside step function. By linearizing the
$k$-dependent spectrum  around $k=k_F$, and exchanging $g$ for the
$(T,v)=(0,0)$ value of the order parameter $\Delta_0$, we arrive at the
following result for $\Delta(T,v)\big|_{T=0}$:
\begin{equation}
\label{eqn:OrderParameter1}
\ln\left(\frac{\Delta}{\Delta_0}\right)=-\int_0^{\omega_D}
d\xi\frac{1}{\sqrt{\xi^2+\Delta^2}}\,\Theta\big(k_F\,
v-\sqrt{\xi^2+\Delta^2}\big).
\end{equation}
This condition yields  two branches of solutions for the order parameter: (1)
If $k_F\, v <\Delta$, we obtain $\Delta=\Delta_0$; on the other hand, (2) if
$k_F\, v
>\Delta$, the condition becomes
\begin{equation}
\ln\left(\frac{\Delta}{\Delta_0}\right)=-\int_0^{\sqrt{k_F^2v^2-\Delta^2}}
\frac{d\xi}{\sqrt{\xi^2+\Delta^2}}=-\sinh^{-1}\left(\frac{\sqrt{k_F^2v^2-\Delta^2}}{\Delta}\right),
\end{equation}
which leads to $\Delta^2=2k_F\, v\, \Delta_0- \Delta_0^2$. These two
zero-temperature solutions were first obtained by Sarma~\cite{Sarma} in
connection with the exchange-field effect in superconductors. In
Fig.~\ref{fig:1} we show these two zero-temperature solutions [i.e., the
horizontal and parabolic curves emanating from the upper right point $(1,1)$],
together with the corresponding solutions at nonzero temperatures. By
continuity, it is not surprising that there is a low-temperature regime in
which there is multivaluedness in the solutions for $\Delta$, as we shall see
in the following subsection and in Fig.~\ref{fig:1}.

\subsection{Nonzero temperatures}
For these, we can solve  Eq.~(\ref{eqn:self_con1}), or equivalently
Eq.~(\ref{eqn:self_con2}), numerically, and obtain the velocity-dependent
order parameter $\Delta(T,v)$ at arbitrary temperatures,   as illustrated  in
Fig.~\ref{fig:1} for several values of the temperature. We remark that in
solving for the order parameter we have linearized the spectrum $E_k\pm kv$
about Fermi momentum $k_F$; this linearization is valid as long as
$\Delta_0/E_F\ll 1$. From Fig.~\ref{fig:1} it is evident that over  a certain
(higher) range of velocities $v$ there are two solutions for $\Delta$, these
solutions differing from those obtained by Bagwell~\cite{Bagwell}, by whom
multiple solutions were not found. This type of multiplicity feature was first
observed by Sarma~\cite{Sarma} in the context of the exchange field in
superconductors. In fact, the self-consistency conditions holding in the
exchange-field case are identical to the ones holding here, provided  one
makes an identification between $k_F\,v$ and the exchange-field energy
$\mu_B\, h$. The two situations appear to be essentially equivalent.
\begin{figure}
 \vspace{0.5cm} \centerline{ \rotatebox{0}{
        \epsfxsize=7.0cm
        \epsfbox{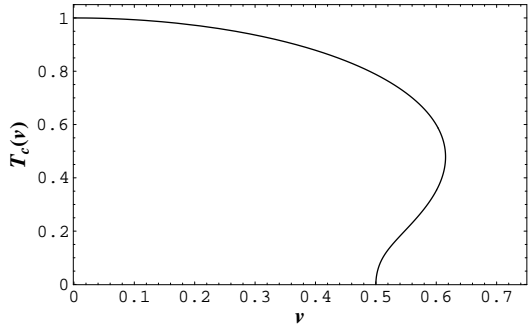}
} }
 \caption{Critical temperature (from linear instability) $T_c(v)$ (in unit of $T_c^0$) vs.~superfluid velocity $v$ (in
 units
 of $v_L$), as obtained from Eq.~(\ref{eqn:TC0}).  Note the occurrence of the unphysical multivalueness.} \label{fig:TC0_V}
\end{figure}

{\changed The feature of double solutions that we have found for the order
parameter, as shown in Fig.~\ref{fig:1}, is in contrast with previous
findings~\cite{Bagwell} (see, in particular, Fig.~3 therein). We suspect that
this discrepancy results from  the use of an iterative scheme for solving the
self-consistency equation numerically;  when multiple solutions exist, such
scheme yields results that depend sensitively on the initial conditions. We
have instead used the bisection method, which locates all possible solutions.
}

\section{Transition temperatures: linear instability}\label{sec:TransitionT}
If one assumes that the transition between the  superconducting and normal
states is continuous, one can determine the transition temperature via the
self-consistency condition taken in  the limit of vanishing $\Delta$. We note
that, owing to the relation of the Fermi function to the Matsubara sum (see,
e.g., Ref.~\cite{Zagoskin}), Eq.~(\ref{eqn:self_con1}) is equivalent to
\begin{eqnarray}
\label{eqn:selfcon3}
 \Delta=g\sum_k T\sum_{\omega_n}\frac{-\Delta}{(i\omega_n-k
v-E_k)(i\omega_n-k v+E_k)},
\end{eqnarray}
where $\omega_n\equiv 2\pi T\big(n+(1/2)\big)$ with $n=0,\pm1,\pm2,\dots,$ are
the Matsubara frequencies. We denote by $T_c$ the value of the  temperature
that solves Eq.~(\ref{eqn:selfcon3})  in the limit $\Delta\rightarrow 0$. In
this limit, imposing a cutoff $\omega_D$ on the Mastubara frequencies, and
integrating out the momentum $k$, we arrive at the condition
\begin{eqnarray}
\label{eqn:Tc} 1=g N_0\,2\pi T_c \sum_{\omega_n>0}^{\omega_D} {\rm
Re}\Big(\frac{1}{\omega_n+i k_F v}\Big),
\end{eqnarray}
in which we have redefined $\omega_n$ to mean  $\omega_n=2\pi
T_c\big(n+(1/2)\big)$. By exchanging $g$ for the zero-flow critical
temperature $T_c^0$ and performing the summation, Eq.~(\ref{eqn:Tc}) becomes
\begin{equation}
\label{eqn:TC0}
 \ln\left(\frac{T_c}{T_c^0}\right)=\psi\Big(\frac{1}{2}\Big)-{\rm
Re}\,\psi\Big(\frac{1}{2}+\frac{i k_F v}{2\pi T_c}\Big),
\end{equation}
where $\psi(x)$ is the di-gamma function~\cite{DiGamma}.

The solutions of Eq.~(\ref{eqn:TC0}) for $T_c$ are shown in
Fig.~\ref{fig:TC0_V}, which, like Fig.~\ref{fig:1}, exhibits a double-solution
feature for a range of velocities. The solution at $T_c=0$ and $k_F\,
v=\Delta_0/2$ corresponds to the branch $\Delta^2=2k_F\, v\, \Delta_0-
\Delta_0^2$. As we shall see below, this branch corresponds to  an {\it
unstable\/} solution; a correct description of the $T_c$ vs.~$v$ phase diagram
requires the   consideration of free energies.

\section{Velocity-temperature phase diagram}\label{sec:Phase}
In the light of the results obtained in Sec.~\ref{sec:TransitionT}, we ask the
question: What is the true transition temperature? To determine this, one has
to take into account the  multiple solutions of the order-parameter
self-consistency condition and compare their free energies (e.g., see
Ref.~\cite{Bardeen}). With this prescription we ask: Will our transition
temperature $T_c(v)$ have the same functional form as the $T_c(h)$ of
Sarma~\cite{Sarma} and of Maki and Tsuneto~\cite{MakiTsuneto} (SMT), in which
$h$ is the magnetic exchange field?

{\changed
 In the case considered by SMT, a comparison is made between the free
energies of a spin-unpolarized superconducting state and a normal state that
is partially spin-polarized, both states being subject to an exchange field.
In the present case the comparison is between a flowing superconducting state
and a stationary normal state. Viewed from a reference frame that flows with
the superconducting state, the normal state flows and can be regarded as the
analog of SMT's polarized state. Furthermore, our superconducting state is the
analog of SMT's superconducting state. This perspective suggests that the
correct procedure for our purposes is to compare the free energies of the
flowing superconducting state and the stationary normal state, as was first
done by Bagwell~\cite{Bagwell}.

The physical reason for the equivalence between the exchange-field case
considered by Sarma and the superflow case considered here is as follows. In
the case of Sarma, due to the exchange field, the up and down electrons in a
Cooper pair have an energy difference of  $\mu_B B$. In the case of flow, the
electron pairing is between states $(k+q)\, \uparrow$ and $(-k+q)\,
\downarrow$, and which have an energy difference of $2k_F q/m$ near the Fermi
surface. In the former case, the normal state can be polarized, and thus can
reduce its energy by an amount proportional to $h^2$. In the latter case, the
stationary normal state has an energy that is lower than the flowing normal
state by an amount proportional to $q^2$. Hence, the two scenarios appear to
be equivalent.

To understand the root of our state-selection procedure, we begin with an
analogy. Consider the liquid and  crystalline states of a particular material.
Following Callen~\cite{callen}, we observe that the information sufficient to
specify uniquely an equilibrium state is greater for the crystalline state
than for the liquid state, owing to the spontaneously broken translational
symmetry of the crystal and the resulting low-energy Goldstone field, i.e.,
the displacement field. Thus, when assessing the relative thermodynamic
stability of a liquid and a crystalline state, one must specify a displacement
field for the crystal but one must not (and indeed cannot) do so for the
liquid state. It is meaningful to speak of an equilibrium crystalline state
that supports a specified static shear stress, but not so for a liquid, as
stress would induce a nonequilibrium dissipative steady state of shear flow,
in which entropy would constantly be being produced. In summary, there is a
unique equilibrium state for the liquid, but there is a family of equilibrium
states for the crystal, differing in their displacement field (or strain
field, or its thermodynamic conjugate, the stress field). One is entitled to
consider the relative thermodynamic stability of a liquid and any one of this
family of crystalline states. Returning now to the case of superconductivity,
the precise analogy is on the one hand between the crystal and the
superconductor, and on the other hand between the liquid and the normal metal.

One is thus entitled to consider the relative thermodynamic stability of a
stationary normal state and any one of the family of the flowing
superconducting states. That there is a family of flowing superconducting
states originates in the spontaneously broken gauge symmetry of the
superconducting state, which amounts to a controllable thermodynamic field,
(i.e., the phase field), which is the analog of the displacement field of the
crystal. (The analogy is: phase  $\Leftrightarrow$ displacement ; velocity
 $\Leftrightarrow$ strain; current density $\Leftrightarrow$ stress.)\thinspace\
By contrast, the normal state is unique.

 To obtain the phase diagram, one should then compare the free energy $F_s(q,\mu)$ of a flowing
superconducting state  to the free energy $F_n(0,\mu)$ of a stationary (i.e.,
not flowing) normal state~\cite{Bagwell}. The energy difference between the
flowing and stationary normal states plays the role of the paramagnetic
energy. We shall see that the true equilibrium transition temperature then has
exactly the functional form obtained by Sarma~\cite{Sarma} and by Maki and
Tsuneto~\cite{MakiTsuneto}.

}

 The free energy~(\ref{eqn:FreeEn}) of the superconducting state at a fixed
chemical potential $\mu$ is given explicitly by
\begin{equation}
\label{eqn:FreeEnS} F_s(q,\mu)=\frac{\Delta^2}{g}+\sum_k
\Big(\frac{k^2}{2m}+\frac{q^2}{2m}-\mu\Big) -T\sum_k \ln\left[2\cosh\beta
\sqrt{\Big(\frac{k^2}{2m}+\frac{q^2}{2m}-\mu\Big)^2+\Delta^2}\,\,+\,\,2\cosh\beta{kv}\right].
\end{equation}
The stationary, normal state at chemical potential $\mu-(q^2/2m)$ thus has
free energy
 \begin{equation}
\label{eqn:FreeEnN0} F_n\Big(0,\mu-\frac{q^2}{2m}\Big)=\sum_k
\Big(\frac{k^2}{2m}+\frac{q^2}{2m}-\mu\Big) -T\sum_k \ln\left[2\cosh\beta
\Big(\frac{k^2}{2m}+\frac{q^2}{2m}-\mu\Big)\,\,+\,\,2\right].
\end{equation}
Our goal is to find the difference between two energies at the same chemical
potential, which is given by
\begin{equation}
\label{eqn:DiffF}
F_s(q,\mu)-F_n(0,\mu)=\left[F_s(q,\mu)-F_n\Big(0,\mu-\frac{q^2}{2m}\Big)\right]-\left[F_n(0,\mu)-F_n\Big(0,\mu-\frac{q^2}{2m}\Big)\right],
\end{equation}
where the term in the second pair of square brackets equals $-N q^2/2m$ (i.e.,
the analog of the paramagnetic energy), with $N$ being the total number of
electrons. Hence, from Eqs.~(\ref{eqn:FreeEnN0}) and~(\ref{eqn:DiffF}) we
arrive at the result that
\begin{eqnarray}
F_s(q,\mu)-F_n(0,\mu)&=& N\frac{q^2}{2m}+\frac{\Delta^2}{g}-T\sum_k
\ln\left[\frac{\cosh\beta
\sqrt{\Big(\frac{k^2}{2m}+\frac{q^2}{2m}-\mu\Big)^2+\Delta^2}\,\,+\,\,\cosh\beta{kv}}{\cosh\beta
\Big(\frac{k^2}{2m}+\frac{q^2}{2m}-\mu\Big)\,\,+\,\,1}\right]\nonumber\\
&=&N\frac{q^2}{2m} + \frac{\Delta^2}{g}-T\sum_k \ln\left[\frac{\cosh\beta
E_k+\cosh\beta{kv}}{\cosh\beta \overline{\veps}_k+1}\right]. \label{eqn:FsFn}
\end{eqnarray}

\begin{figure}
 \vspace{0.5cm} \centerline{ \rotatebox{0}{
        \epsfxsize=7.0cm
        \epsfbox{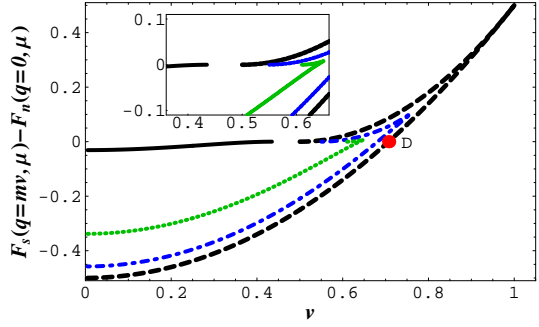}
} }
 \caption{Free-energy difference {\changed $F_s(q=mv,\mu)-F_n(0,\mu)=Nq^2/2m-\delta F$ (in units of $N_0 \Delta_0^2$) at various
 temperatures $T=(0, 0.2,0.4,0.85)\, T_c^0$ (black dashed, blue dash-dotted, green dotted, black solid, respectively) vs. superfluid velocity $v$ (in units of
$v_L$). When $F_s(q,\mu)-F_n(0,\mu)$ is negative, the corresponding
superconducting solution is stable. The red dot (D) indicates the transition
point at $v_c=v_L/\sqrt{2}$ and $T_c=0$; see also the point D in
Fig.~\ref{fig:sfPhase}. The two branches for lower temperatures ($T=0, 0.2,
0.4$) represent double solutions for the order parameter. The range of $v$ for
which a nonzero solution exists shrinks as the temperature increases, and the
feature of double solutions disappears at sufficiently high temperatures; see
also Fig.~\ref{fig:1}. The inset shows a blowup in the range $[0.35,0.7]$ for
$v$.} } \label{fig:FreeEn}
\end{figure}
What we should do to obtain the phase diagram is to compare the possible
nonzero solution(s) for the order parameter to determine which solution
 possesses the lower free energy $F_s(q,\mu)$ and whether this
free energy is lower than that of a stationary (i.e., nonmoving) normal state,
i.e., whether $F_s(q,\mu)-F_n(0,\mu){\changed \le} 0$. For convenience, let us
make the definition
\begin{equation}
\label{eqn:dFq2} \delta F(v,\mu)\equiv N\frac{q^2}{2m}
-F_s(q,\mu)+F_n(0,\mu)=-\frac{\Delta^2}{g}+T\sum_k \ln\left[\frac{\cosh\beta
E_k+\cosh\beta{kv}}{\cosh\beta \overline{\veps}_k+1}\right].
\end{equation}
 The solution for $\Delta$ that has the largest value of $\delta F-
N{q^2}/{2m}$, provided this difference is greater than zero, is the true
equilibrium solution; otherwise the normal state will be the true equilibrium
state.
 The kinetic flow term $N{q^2}/{2m}$ can be {\changed re-expressed} (by relating the total electron number
 to the density of state) {\changed as} $N_0\, \Delta_0^2\, (v/v_L)^2$, where
$N_0$ is the density of states per spin, and we have, for convenience, set the
total ``volume'' of the system to be unity. Note that from
Eq.~(\ref{eqn:dFq2})  we have at zero temperature
 \begin{eqnarray}
\label{eqn:FzeroT} \delta
F(v,\mu)=N_0\,\frac{\Delta^2}{2}+2\sum_{k}\Big(kv-E_k+\frac{\Delta^2}{2E_k}\Big)\Theta(kv-E_k).
 \end{eqnarray}
 This expression also holds for higher-dimensional systems, provided we
  replace $k$ by $\vec{k}$, $v$ by $\vec{v}$, and $kv$ by $\vec{k}\cdot\vec{v}$.
The effective condensation energy $\delta F$ for the solution
$\Delta=\Delta_0$ [see the discussion below Eq.~(\ref{eqn:OrderParameter1})]
 is $N_0 \,\Delta_0^2/2$. Balancing $\delta F$ against the additional energy $Nq^2/2m=N_0\, \Delta_0^2\,
(v/v_L)^2$ due to the flow, we obtain $v_c/v_L= 1/\sqrt{2}$. This corresponds
to the Clogston-Chandrasekhar limit in the exchange-field
case~\cite{ClogstonChandrasekhar}.

\begin{figure}

 \vspace{0.5cm} \centerline{ \rotatebox{0}{
        \epsfxsize=7.0cm
        \epsfbox{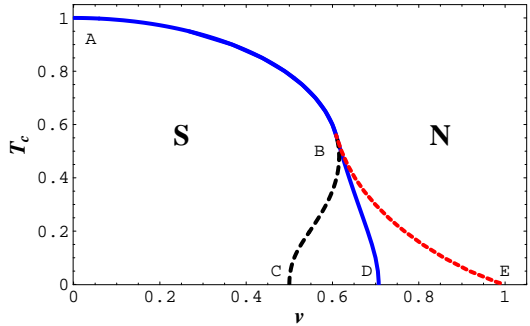}
} }
 \caption{Temperature versus flow-velocity phase diagram, indicating normal (N) and superconducting (S)
 regions. The phase boundary ABD indicates the transition between superconducting and normal
 equilibrium states, {\changed which is obtained using the principle described below Eq.~(\ref{eqn:FsFn})}. Across segment AB of the phase boundary the superconducting-normal transition
 is continuous. Across segment BD the transition is discontinuous. As discussed in Sec.~\ref{sec:Super},
 line segments BC and BE indicate linear stability limits for, respectively, the normal and superconducting
 metastable states.  $T_c$ and $v$ are measured in units of $T_c^0$ (i.e., the zero-flow critical temperature) and $v_L$, respectively.
  } \label{fig:sfPhase}
\end{figure}
In Fig.~\ref{fig:FreeEn}, we plot $\delta F$ versus the momentum $q$ for the
two superconducting solutions at several temperatures, as well as $N q^2/2m$.
The crossing  of the curve representing $N q^2/2m$ by the curve representing
the superconducting state signifies a discontinuous transition. By finding
such transitions  at various values of $v$, we arrive at the equilibrium phase
boundary $T_c(v)$, shown as the curve ABD in Fig.~\ref{fig:sfPhase}.

\section{Supercooling and superheating}\label{sec:Super}
As discussed in Sec.~\ref{sec:Phase}, we have discussed the true equilibrium
phase transition curve ABD; see Fig.~\ref{fig:sfPhase}. But what is the
physical meaning of the branch BC? {\changed It would be a limit of
metastability if a normal state with flow were a true equilibrium state.
Suppose that the system were in an equilibrium normal state with $v>v_c$
(i.e., lying to the right of the equilibrium phase boundary ABD), and were
then `quenched' by the rapid change of $T$ and/or $v$ into the metastable
region BCD. The system would remain in the normal state, but metastably so,
until a fluctuation were to occur that would nucleate a droplet of the
superconducting phase large enough to grow and complete the conversion of the
state to the true equilibrium state, i.e., the superconducting state. The
metastability limit (which is also known as a spinodal line) is formally
obtained by solving for the temperature obeying the equation
[c.f.~Eqs.~(\ref{eqn:self_con1}),~(\ref{eqn:selfcon3}) and~(\ref{eqn:Tc})]
\begin{equation}
 \lim_{\Delta\rightarrow 0}\frac{1}{\Delta}\frac{\delta F}{\delta
\Delta}=0.
\end{equation}
However, as we argue in Sec.~\ref{sec:Phase}, a normal state with a constant
flow is not a true equilibrium state; therefore, strictly speaking, the curve
BC does not represent a meaningful supercooling curve.

 On the other hand, the
branch BE is a true limit of metastability for the superconducting state. }
The system can be `quenched' into a metastable superconducting state from an
equilibrium superconducting state [i.e., from $(T,v)$ to the left of the
equilibrium phase boundary] by rapidly changing $(T,v)$ to a value in the
region BED. The system will remain superconducting, but metastably so, until a
normal-phase nucleation event carries it to the equilibrium (i.e., normal)
state. In this case, the metastability limit is determined by simultaneously
solving the equations $\delta F/\delta \Delta=0$ and $\delta^2 F/\delta^2
\Delta=0$ (but $\Delta\ne0$) numerically; see Fig.~\ref{fig:sfPhase}.

We note that a corresponding diagram in the case of exchange-field effect in
superconductors was first obtained by Maki and Tsuneto~\cite{MakiTsuneto}. To
answer the question of  how long it would take for system in a metastable
state to find its equilibrium state is a kinetic one that would require
further investigation.

\section{Supercurrent}\label{sec:current}
In this section we examine the dependence of the supercurrent on the flow
velocity, and thereby determine the critical current, i.e., the maximum
equilibrium supercurrent that the system can sustain at various temperatures.
In general, the charge current $I_Q$ carried by the system is defined via
\begin{equation}
\label{eqn:IQ} I_Q\equiv e\sum_{k,\sigma} \frac{q+k}{m}\langle
c_{k+q,\sigma}^\dagger\, c_{k+q,\sigma} \rangle.
\end{equation}
As $\sum_{k,\sigma}\langle  c_{k+q,\sigma}^\dagger c_{k+q,\sigma} \rangle$ is
the total number of electrons $N$, we can re-write Eq.~(\ref{eqn:IQ}) as
\begin{equation}
I_Q= \frac{e}{m}N\,v+e\sum_{k,\sigma} \frac{k}{m}\langle
c_{k+q,\sigma}^\dagger\, c_{k+q,\sigma} \rangle,
\end{equation}
where $v\equiv q/m$.
 By using Eqs.~(\ref{eqn:BogoVale}) and~(\ref{eqn:f12}), the second term can
 be simplified to $(2e/m)\sum_k k\, f_{1;k}$, and thus the expression for $I_Q$
 becomes
 \begin{equation}
 \label{eqn:IQ2}
 I_Q(T,v)=-\frac{|e|}{m}\Big\{N\,m\,v+2\sum_k k f_{1;k}\Big\},
 \end{equation}
 where the electron charge is $e=-|e|$ and the temperature $T$ is implicit in the Fermi function $f_{1;k}$; see
 Eq.~(\ref{eqn:f12}).

\begin{figure}
\vspace{0.5cm} \centerline{ \rotatebox{0}{
        \epsfxsize=7.0cm
        \epsfbox{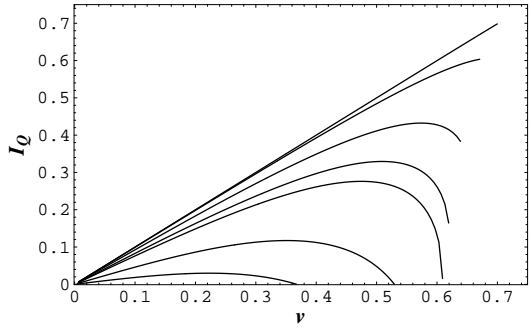}
} }
 \caption{Supercurrent $I_Q$ (in units of $I_Q^0$) vs.~superfluid velocity $v$ (in unit of $v_L$) for various temperatures.
 From top to bottom: $T=(0.1,0.25,0.4,0.5,0.556,0.75,0.9)\, T_c^0$. The curves
 terminate at the  critical velocities $v_{\rm c}(T)$ appropriate to these temperatures. The maximum
 supercurrent  for a
 particular curve determines  the value of the critical current at that temperature. } \label{fig:IvsV}
\end{figure}
We shall compute the current in the superconducting state as a function of $T$
and $v$, and then, by choosing the value $v_m(T)$ of $v$ in the range
$[0,v_c(T)]$ that maximizes $I_Q(T,v)$ at fixed $T$, we shall obtain the
critical current, denoted by $I_C(T) \big[\equiv I_Q\big(T,v_m(T)\big)\big]$,
at that temperature. Equation~(\ref{eqn:IQ2}), applied at $T=0$, tells us that
the zero-temperature current is given by
\begin{eqnarray}
I_Q(T=0,v)&=&-\frac{|e|}{m}\Big[N\,m\,v-2N_0\, k_F
\sqrt{k_F^2v^2-\Delta_0^2}\,\Theta(k_F\,
v-\Delta_0)\Big],\\
&=&I_Q^0\left[\frac{v}{v_L} -
\sqrt{\frac{v^2}{v_L^2}-1}\,\,\Theta(v-v_L)\right],
\end{eqnarray}
where $I_Q^0\equiv -N|e|\Delta_0/k_F=-2|e|\Delta_0/\pi$. Although, formally,
$I_Q(T=0,v)$ achieves its maximum value (viz., $I_Q^0$) at $k_F \,v=\Delta_0$,
this maximum value is unattainable, because the superconducting state gives
way to the normal state via a pre-emptive transition at $k_F\,
v=\Delta_0/\sqrt{2}$ (unless the system falls out of equilibrium and remains
metabstably in the superconducting state). Therefore, the maximum equilibrium
critical current at zero temperature occurs at $v=v_L/\sqrt{2}$, and has the
value $I_Q^0/\sqrt{2}$. At $v=v_L/\sqrt{2}$ the superconducting state first
becomes unstable, so the
 critical current that is attainable in practice is slightly less than $I_Q^0/\sqrt{2}$. This is
unlike the behavior in three dimensions, as well as the higher of the nonzero
temperatures; for these situations the maximum current is achieved at a value
of superfluid velocity for which the superconducting state is still globally
stable.

\begin{figure}
 \vspace{0.5cm} \centerline{
\rotatebox{0}{
        \epsfxsize=7.0cm
        \epsfbox{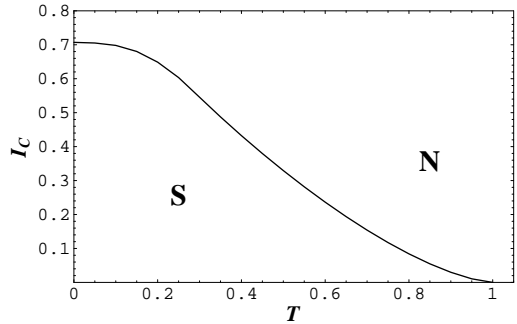}
} }
 \caption{Critical current $I_c$ (in units of $I_Q^0$) vs.~temperature $T$ (in units of
 $T_c^0$). This is also current vs.~temperature phase diagram.
  } \label{fig:IcVsT}
\end{figure}
In Fig.~\ref{fig:IvsV} we show the equilibrium supercurrent $I_Q$ vs. the
superfluid velocity~$v$ for various values of the temperature. The value of
$v=v_m$ that corresponds to a maximum in the current (i.e., the critical
current $I_c$) is not necessarily $v_{\rm c}$. (It {\it is\/} $v_c$ for lower
temperatures, but not for higher ones.) The temperature dependence of the
critical current $I_c(T)$ is shown in Fig.~\ref{fig:IcVsT}.

\begin{figure}
 \vspace{0.5cm} \centerline{ \rotatebox{0}{
        \epsfxsize=7.0cm
        \epsfbox{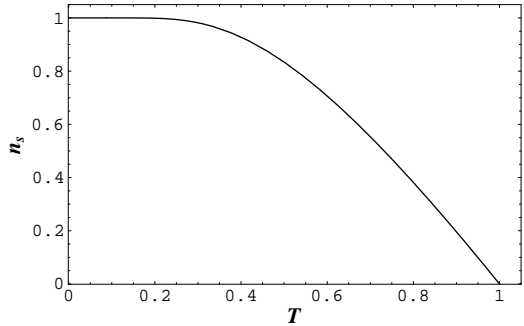}
} }
 \caption{Superfluid density $n_s(T)$ (in units of $2e k_F/\pi$) vs.~temperature $T$ (in units of $T_c^0$).
  } \label{fig:nsVsT}
\end{figure}
A related quantity of interest is the superfluid density $n_s$, defined as
\begin{equation}
n_s=\frac{\partial j_s(v)}{\partial v}\Big|_{v=0},
\end{equation}
i.e.,  the response of the current density $j_s$ (whose expression is
identical to $I_Q$ because we have set the system `volume' to unity) to an
infinitesimal change in the `driving' velocity $v$. The temperature dependence
of $n_s$ is shown in Fig.~\ref{fig:nsVsT}. The vanishing of $n_s$ near $T_c$
is, as expected, linear, and the departure of $n_s$ from its zero-temperature
value has, at low temperatures, a thermally activated form.

\section{Effect of disorder}\label{sec:Disorder}
So far, we have discussed clean systems, i.e., systems not disordered by any
scattering electrons by impurities. What effect will disorder have? In
particular, would the disorder change the nature of the
superconducting-to-normal transition? In this section we briefly discuss the
latter issue. In the case of disorder, the Green function technique is better
suited for dealing with disorder than is trying to diagonalize the Hamiltonian
with inclusion of an arbitrary configuration of the impurities. An equivalent
calculation in the context of the exchange field was done by Maki and
Tsuneto~\cite{MakiTsuneto} and, more recently, by Wei and
Goldbart~\cite{WeiGoldbart} in the context of Little-Parks effect in small
rings. Here, in contrast with Ref.~\cite{WeiGoldbart}, we are not concerned
with the finiteness of the system size, and we simply quote the relevant
results for the equation obeyed by the critical temperature, i.e., Eq.~(B39)
of Ref.~\cite{WeiGoldbart}:
\begin{eqnarray}
\label{eqn:arbitraryDisorder}
\ln\left(\frac{T_\text{c}(v)}{T_\text{c}^0}\right)
&=&\psi\left(\frac{1}{2}\right)-\frac{1}{\sqrt{\alpha^2-\chi^2}}
\left[\frac{-\alpha+\sqrt{\alpha^2-\chi^2}}{2}\,\psi\left(\frac{1+\alpha+\sqrt{\alpha^2-\chi^2}}{2}\right)
\right.\nonumber \\
&&\qquad\left.+\,\frac{\alpha+\sqrt{\alpha^2-\chi^2}}{2}\,\psi\left(\frac{1+\alpha-\sqrt{\alpha^2-\chi^2}}{2}\right)
\right],
\end{eqnarray}
where, in the present case, $\alpha\equiv 1/4\pi\tau_0\,T_\text{c}(v)$ and
$\chi\equiv k_F\, v/\pi\, T_\text{c}(v)$, the parameter $\tau_0$ is the
elastic mean-free time, and $\psi(x)$ is the di-gamma function~\cite{DiGamma}.

\begin{figure}
 \vspace{0.5cm} \centerline{ \rotatebox{0}{
        \epsfxsize=7.0cm
        \epsfbox{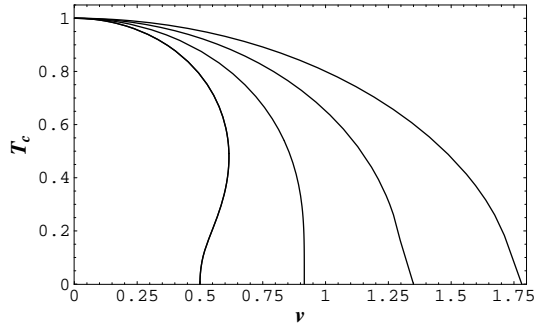}
} }
 \caption{The impact of elastic scattering on the flow-velocity--dependent normal-to-superconducting
 transition temperature solutions obtained via the self-consistency, Eq.~(\ref{eqn:arbitraryDisorder}).  $T_c$ and $v$ are in units of $T_c^0$ and $v_L$, respectively.
 From left to right: $\hbar/\tau_0\,\Delta_0=(0,1,1/0.546,5,10)$, i.e., cleaner to dirtier.}
\label{fig:TvsVDisorder}
\end{figure}
In the clean limit (i.e., $\tau_0\,\Delta_0\gg 1$),
Eq.~(\ref{eqn:arbitraryDisorder}) reduces to  Eq.~(\ref{eqn:TC0}), for which
multiple solutions for $T_c(v)$ exist. For strong disorder
($\tau_0\,\Delta_0\ll 1$), Eq.~(\ref{eqn:arbitraryDisorder}) reduces to
\begin{equation}
\label{eqn:TcDisorder}
 \ln\left(\frac{T_\text{c}(v)}{T_\text{c}^0}\right)
=\psi\left(\frac{1}{2}\right)-\psi\left(\frac{1}{2}+\frac{\chi^2}{4\alpha}\right),
\end{equation}
for which no multiple solutions for $T_c(v)$ exist. This suggests that the
assumption of a vanishing order parameter in the search for the critical
temperature is valid and, hence, that the transition is continuous. Moreover,
from Eq.~(\ref{eqn:TcDisorder}), in the strong-disorder limit the critical
velocity is determined via
\begin{equation} \label{eqn:vcdisorder}
k_F\, v_{c}/\Delta_0=1/\sqrt{4\tau_0\,\Delta_0}. \end{equation} Numerically,
we have found  that  for $\tau_0\,\Delta_0\lesssim 0.55$ multiple solutions do
not occur. This, then, gives the threshold for the disorder strength that
divides the discontinuous and continuous transition regimes. In
Fig.~\ref{fig:TvsVDisorder} we show the solutions of
Eq.~(\ref{eqn:arbitraryDisorder}) for various values of $\tau_0\,\Delta_0$. In
the strong-disorder limit we obtain from Eq.~(\ref{eqn:vcdisorder}) that
$v_c/v_L=\sqrt{\pi \xi_0/4\l_e}$ (where $l_e$ is the elastic mean-free path
and $\xi_0$ is zero-temperature coherence length), in particular that  the
critical velocity exceeds $v_L$. But will the critical {\it current\/} also
exceed its clean-limit value? According to Bardeen~\cite{Bardeen}, the
superfluid density in the strong-disorder limit is reduced by a factor of
roughly $l_e/\xi_0$. An order-of-magnitude estimate gives for the critical
current the product of the superfluid density $n_s\sim (l_e/\xi_0) n_s^0$
(where $n_s^0$ denotes the superfluid density in the clean limit) and the
critical velocity $v_c=\sqrt{\pi \xi_0/4\l_e}v_L$. This, in turn, gives that
the critical supercurrent is reduced  by roughly a factor $\sqrt{l_e/\xi_0}$,
which is much less than unity in the strong-disorder limit. Therefore, even
though the critical velocity is increased by disorder, the critical current is
reduced by it.

\begin{figure}

\vspace{0.5cm} \centerline{ \rotatebox{0}{
        \epsfxsize=7.0cm
        \epsfbox{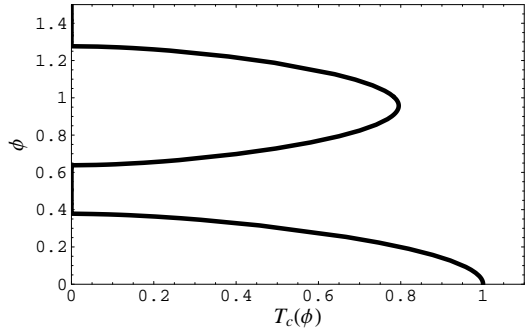}
} }
 \caption{
 {\changed Critical temperature $T_c(\phi)$ (in unit of zero-flux critical temperature $T_c^0$) vs. magnetic flux
 [$\phi\equiv \Phi/(hc/2e)$] for a superconducting ring. The result is
 obtained by using Eq.~(\ref{eqn:bulk_dirty}) with the additional pair-breaking
 term given in Eq.~(\ref{eqn:thickness}), and a numerical factor of about $2.5$
  is the only fitting parameter. All other parameters
 are taken from those reported in the experiment by Liu et al.~\cite{PennState}. }
 }
\label{fig:Liu}
\end{figure}

{\changed
 We now digress to discuss the scenario of a closed superconducting loop threaded
by magnetic flux, which is pertinent to the experiment performed by Liu et
al.~\cite{PennState} and the situation considered theoretically in
Ref.~\cite{WeiGoldbart}. There, the boosted velocity is induced by the
magnetic flux. The measurement of $T_c$ vs. flux $\phi\equiv \Phi/\Phi_0$
(with $\Phi_0\equiv hc/2e$) presented in Fig.~4 of Ref.~\cite{PennState} shows
non-$\Phi_0$-periodic behavior, with the first dome (near $\phi=0$) being
higher than the second dome (near $\phi=1$). Naturally, the data explore only
a limited range of flux, so it is not possible to determine with certainty
whether higher flux values would yield dome heights that oscillate with flux
or decay. Oscillations would be a consequence of flux through the hole; decay
would be a consequence of flux through the sample, which would cause pair
breaking. To ascertain which of these two effects (oscillation or decay) is
likely to dominate in the range of fluxes probed by the experiment, it is
useful to compare several lengthscales. First, the sample radius $R\approx
75\,{\rm nm}$ is smaller than the coherence length $\xi_0$, by a factor of
roughly $0.06$. Following Ref.~\cite{WeiGoldbart}, this suggests the smallness
of the radius as the origin of the dome-height mismatch. On the other hand, a
more complete characterization of the system would take into account the
elastic mean-free path $l_e$, which for the sample studied in
Ref.~\cite{PennState} is shorter than the radius, by a factor of approximately
$7.8$. This puts the sample in the dirty regime, and as a result, the impact
of the smallness of the radius is suppressed and tends to restore the $\Phi_0$
periodicity of $T_c$ discussed in Ref.~\cite{WeiGoldbart} and given by
\begin{equation}
\label{eqn:bulk_dirty} \ln\left(\frac{T_c(\phi)}{T_c^0}\right)=\ln
t(\phi)=\psi\Big(\frac{1}{2}\Big)-\psi\Big(\frac{1}{2}+\frac{\Gamma
 l_e\xi_0x_m^2(\phi)}{t(\phi) R^2}\Big),
\end{equation}
where $\Gamma\approx1.76$ and $x_m(\phi)=\min_{m\in Z} |\phi -m|/2$. What else
could cause the lack of $\Phi_0$ periodicity? Because the magnetic flux is not
confined inside the hole in the loop, but is also present throughout the
sample, owing to its finite thickness, the orbital pair-breaking effect comes
into play. To account for this effect, we use the scenario of the orbital
pair-breaking effect of a parallel magnetic field applied to a film of finite
thickness $d$ (which is about $25\,{\rm nm}$ in the experiment), which leads
us to add to the argument of the second term on the r.h.s.~of
Eq.~(\ref{eqn:TcDisorder}) the term
\begin{equation}
\label{eqn:thickness} \frac{1}{6}\frac{D e^2 B^2 d^2}{2\pi k_B T_c(\phi)\hbar
c^2}=\frac{\Gamma \phi^2 \xi_0 l_e d^2}{36\, t(\phi) R^4};
\end{equation}
see, e.g., Ref.~\cite{Tinkham}. (The term should be correct up to a numerical
factor due to geometry.)\thinspace\ As we see from Fig.~\ref{fig:Liu}, the
resulting dependence of the critical temperature on the flux $\phi$ seems to
reproduce, at least qualitatively, the features reported by Liu et
al.~\cite{PennState}. }

\section{Higher dimensions}\label{sec:higherD}

We have seen that in the clean limit the discontinuous transition at flow
velocity $v=\Delta_0/k_F \sqrt{2}$ occurs due to a competition between the
condensation energy $N_0\,\Delta_0^2/2$ and the flow kinetic energy $N
q^2/2m=N_0\, \Delta_0^2\, (v/v_L)^2$ (for one dimension; see
footnote~\cite{footnoteDOS}), where $v_L\equiv \Delta_0/k_F$. {\changed The
principle that we employed in Sec.~\ref{sec:Phase} is to compare the free
energies of the flowing superconducting states and of the stationary normal
state. As further confirmation of the validity of this principle, we now apply
it to higher-dimensional settings, and compare the results with the known one
mentioned in the Introduction.}
\begin{figure}

\vspace{0.5cm} \centerline{ \rotatebox{0}{
        \epsfxsize=7.0cm
        \epsfbox{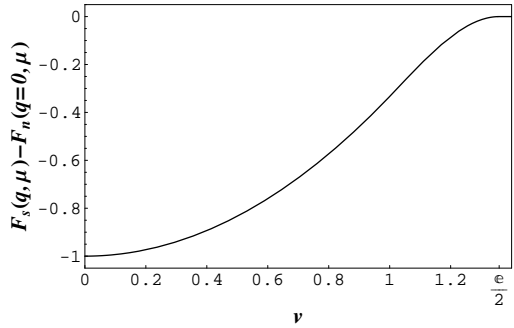}
} }
 \caption{The energy difference {\changed $F_s(q=mv,\mu)- F_n(0,\mu)$ }(in units of $N_0\,\Delta_0^2/2$) versus~flow velocity $v$ (in units of $v_L$) in three spatial dimensions.
 When the difference is less than zero, the flowing superconducting state is stable.}
\label{fig:dFq2Vsv}
\end{figure}

 In two dimensions, the solution of the self-consistency
equation~(\ref{eqn:self_con2}) gives  $\Delta=\Delta_0$ for $v\le v_L$, and
$\Delta=0$ for $v>v_L$. This indicates a discontinuous transition at $v_c^{\rm
(2D)}=v_L$. Indeed, this result is consistent with free-energy considerations:
by using the condensation energy $\delta F= N_0\,\Delta_0^2/2$ and the flow
kinetic energy $N q^2/2m=N_0 \Delta_0^2(v/v_L)^2/2$ (the difference  from the
one-dimensional case arising from the  density of states; see
footnote~\cite{footnoteDOS}), we conclude that $v_c^{\rm (2D)}=v_L$.

In three dimensions the solution of the self-consistency
equation~(\ref{eqn:self_con2}) gives $\Delta=\Delta_0$ for $v<v_L$, and for
$v_L\le v\le (e/2)\,v_L$ (see Ref.~\cite{Zagoskin}) gives the following
implicit equation for $\Delta$:
\begin{eqnarray}
\label{eqn:3DDv} \sqrt{1-\Big(\frac{\Delta}{k_F\, v}\Big)^2}-\ln
\left[1+\sqrt{1-\Big(\frac{\Delta}{k_F\, v}\Big)^2}\,\right]=\ln \left(\frac{
v}{v_L}\right),
\end{eqnarray}
where we have ignored terms of order $\Delta/\omega_D$ or smaller (with
$\omega_D$ denoting the Debye frequency). Equation~(\ref{eqn:3DDv}) gives that
$\Delta=0$ at $v=v_c^{\rm (3D)}=(e/2)\,v_L\approx 1.359\, v_L$. Is this
linear-instability result consistent with free-energy considerations (i.e.,
the balancing of the two energies)? If one were to na{\"\i}vely use the
condensation energy $N_0\, \Delta_0^2/2$ to balance the flow energy $N q^2/2m=
N_0 \,\Delta_0^2\, ( v/v_L)^2/3$~\cite{footnoteDOS}, one would obtain
$v_c=\sqrt{3/2}\,\,(\approx 1.225)\,v_L< v_c^{\rm (3D)}$! Does this mean that
the superconducting state really becomes unstable at $v\approx 1.225\, v_L <
v_c^{\rm (3D)}$? Such a result would be in conflict with the result, mentioned
in Sec.~\ref{sec:Intro}, that $v_c ^{\rm (3D)}= (e/2)\, v_L$? How is this
apparent conflict resolved? For $v\le v_L$, the effective condensation energy
$\delta F$ is indeed $ N_0 \,\Delta_0^2/2$. But for $v>v_L$, $\delta F \ne N_0
\,\Delta_0^2/2$, so it would be incorrect to equate $ N_0 \,\Delta_0^2/2$ and
$Nq^2/2m$ in order to deduce $v_c$. Instead, one should use the expression for
$\delta F$ appropriate to the range $v>v_L$. In fact, an evaluation of
Eq.~(\ref{eqn:FzeroT}) in three dimensions gives (for $v>v_L$)
\begin{eqnarray}
\label{eqn:dF3D} \delta F&=&\frac{N_0}{2}\Delta^2+\frac{N_0}{3}(k_F
v)^2\left[\sqrt{1-\Big(\frac{\Delta}{k_F v}\Big)^2}\,\right]^3 \nonumber\\
&&+N_0\, \Delta^2 \sinh^{-1}\frac{\sqrt{(k_F
v)^2-{\Delta}^2}}{\Delta}-N_0\,\Delta^2\ln\frac{\sqrt{(k_F\,
v)^2-{\Delta}^2}+k_F\, v}{\Delta},
\end{eqnarray}
where the dependence of $\Delta$ on $v$ is given by Eq.~(\ref{eqn:3DDv}). From
Eqs.~(\ref{eqn:FzeroT}) and~(\ref{eqn:dF3D}) we have that for
$v_L<v<(e/2)\,v_L$,
 $\Delta> 0$ and $\delta F -N q^2/(2m) > 0$. Only when $v$ increases to $v=(e/2)\,v_L$ (i.e.,
when $\Delta=0$) do we have $\delta F -(N q^2/2m)= 0$. Thus, the transition at
$v=(e/2)\, v_L$ is continuous, as the order parameter vanishes continuously
there, and, hence, the conflict is resolved. The dependence of the quantity
$F_s(q,\mu)-F_n(0,\mu)=N q^2/(2m)-\delta F$ on  flow velocity $v$ is plotted
in Fig.~\ref{fig:dFq2Vsv}, which clearly shows that the superconducting state
is stable for flow velocities in the range $0\le v<(e/2)\,v_L$.

\section{Concluding remarks}\label{sec:Conclude}
We have revisited the issue of the critical velocity of an effectively
one-dimensional superconductor {\changed using mean-field theory}. As we have
discussed, this issue is equivalent to that of the threshold uniform
exchange-field in a superconductor, first investigated by Sarma~\cite{Sarma}
and by Maki and Tsuneto~\cite{MakiTsuneto}. {\changed This equivalence is due
to the correspondence of (i)~the Zeeman frustration energy $\mu_B B$ between
the two electrons of a Cooper pair (in the case of exchange field), and
(ii)~the kinetic frustration energy $2k_F v$ (in the case of flow). In
particular, we have found that at zero temperature the critical velocity is a
factor of $\sqrt{2}$ smaller than the Landau critical velocity, in contrast to
a previous finding~\cite{Bagwell}. Our result originates in  a
Clogston-Chandrasekhar--like discontinuous transition between the
superconducting and the normal state.
 The physical reason for the discontinuity of the transition is that it results
 from a balance between the condensation and  flow energies. (Recall that
  the bulk superconducting phase transition in magnetic field results from
  a balance between the
condensation ($N_0 \Delta_0^2/2$) and  magnetic field ($B^2/8\pi$) energies,
and that the transition is discontinuous.)\thinspace\ The transition remains
discontinuous for temperatures below approximately $0.56 T^0_c$; above that,
it is continuous.

 We have also
studied the issue of critical supercurrents, determined the
current-temperature phase diagram, and examined metastability (and its limits)
in the temperature versus flow-velocity phase diagram.  As a test of our
underlying principle, namely the comparison of the free energies of flowing
superconducting and stationary normal states, we have also examined the two-
and three-dimensional cases, and have thus obtained results that are in
agreement with previous findings. } In addition, we have commented on the
effects of electron scattering by impurities and, in particular, we have
argued that strong disorder renders continuous the aforementioned
discontinuous phase transition. {\changed Even though disorder can increase
the value of the critical {\it velocity\/}, the physically measurable
quantity, i.e., the critical {\it current\/}, is still reduced by the
disorder. }

{\changed Throughout this Paper, we have adopted a mean-field approach, thus
ignoring the effects of fluctuations such as phase slips. Furthermore, we have
limited ourselves to the analysis of equilibrium properties as well as issues
of metastability. The intriguing issue of how a flowing superconducting state
undergoes the transition to the stationary normal state, which essential
dynamical processes take place, and on what timescale all remain as research
directions for the future. }

\medskip \noindent {\it Acknowledgments\/}. The authors acknowledge useful
discussions with Anton Burkov, Tony Leggett and Frank Wilhelm. This work was
supported by IQC, NSERC, and ORF (TCW) and by DOE Grant No.~DEFG02-07ER46453
through the Frederick Seitz Materials Research Laboratory at the University of
Illinois (PMG).


\begin{thebibliography}{99}
\bibitem{Landau}
L. D. Landau, J. Phys. USSR {\bf 5}, 71 (1941).
\bibitem{Varoquaux06}
E. Varoquaux, C. R. Physique {\bf 7}, 1101 (2006).
\bibitem{Rogers}
K. T. Rogers, Ph.D. thesis, University of Illinois, 1960.
\bibitem{Bardeen}
J. Bardeen, Rev. Mod. Phys. {\bf 34}, 667 (1962).
\bibitem{Zagoskin} For example, see A. M. Zagoskin,
 {\it Quantum Theory of Many-Body Systems\/}
(Springer-Verlag, New York, 1998).
\bibitem{Bagwell} P. F. Bagwell, Phys. Rev. B {\bf 49}, 6841 (1994).
\bibitem{ClogstonChandrasekhar}
A. M. Clogston, Phys. Rev. Lett. {\bf 9}, 266 (1962); B. S. Chandrasekhar,
Appl. Phys. Lett. {\bf 1}, 7 (1962).
\bibitem{Sarma}
 G. Sarma, J. Phys. Chem. Solids, {\bf 24}, 1029 (1963).
\bibitem{MakiTsuneto}
K. Maki and T. Tsuneto, Prog. Theor. Phys. {\bf 31}, 945 (1964).
\bibitem{KetterleGroup}
C. Raman, M. K\"ohl, R. Onofrio, D. S. Durfee, C. E. Kuklewicz, Z. Hadzibabic,
and W. Ketterle, Phys. Rev. Lett. {\bf 83}, 2502 (1999); D. E. Miller, J. K.
Chin, C. A. Stan, Y. Liu, W. Setiawan, C. Sanner, and W. Ketterle, Phys. Rev.
Lett. {\bf 99}, 070402 (2007).
\bibitem{LangerFisher67}
J. S. Langer and M. E. Fisher, Phys. Rev. Lett. {\bf 19}, 560 (1967).
\bibitem{LA}
J. S. Langer and V. Ambegaokar, Phys. Rev. {\bf 164}, 498 (1967).
\bibitem{QPS}
A. D. Zaikin, D. S. Golubev, A. van Otterlo, and G. T. Zimanyi,  Phys. Rev.
Lett. {\bf 78}, 1552 (1997).
\bibitem{NozieresSchmitt-Rink}
P. Nozi\`eres and S. Schmitt-Rink, J. Low Temp. Phys. {\bf 59}, 195 (1985).
\bibitem{BCS}
J. Bardeen, L. N. Cooper, and J. R. Schrieffer, Phys. Rev. {\bf 106}, 162
(1957).
\bibitem{DiGamma} M. Abramowitz and I. A. Stegun, {\it Handbook of Mathematical
Functions\/}, (Dover, New York, 1972), pp. 259 and pp. 556.
\bibitem{callen}
{\changed
 H. B. Callen, {\it Thermodynamics and an Introduction to
Thermostatistics\/}, (John Wiley, New York, 1985). }
\bibitem{WeiGoldbart}
T.-C. Wei and P. M. Goldbart, Phys. Rev. B {\bf 77}, 224512 (2008).
\bibitem{PennState}
 {\changed
Y. Liu, Yu. Zadorozhny, M. M. Rosario, B. Y. Rock, P. T. Carrigan, and H.
Wang, Science {\bf 294}, 2332 (2001).}
\bibitem{Tinkham}
{\changed M. Tinkham, {\it Introduction to Superconductivity\/}, (McGraw-Hill,
New York, 1996), p.393. }
\bibitem{footnoteDOS}
Because the expression for the density of states depends on spatial dimension,
we have that: (a) in one dimension, $N q^2/2m=N_0\, \Delta_0^2\, (v/v_L)^2$;
(b) in two dimensions, $N q^2/2m=N_0\, \Delta_0^2\, (v/v_L)^2/2$; and (c) in
three dimensions, $N q^2/2m=N_0\, \Delta_0^2\, (v/v_L)^2/3$.
\end{thebibliography}
\end{document}